# Nanoscale Optical Imaging of 2D Semiconductor Stacking Orders by Exciton-Enhanced Second Harmonic Generation

Kaiyuan Yao[1], Shuai Zhang[2], Emanuil Yanev[1], Kathleen McCreary[3], Hsun-Jen Chuang[3,4], Matthew R. Rosenberger[5], Thomas Darlington[1], Andrey Krayev[6], Berend T. Jonker[3], James C. Hone[1], D.N. Basov[2], P. James Schuck[1]

[1] Department of Mechanical Engineering, Columbia University, New York, NY, USA;

[2] Department of Physics, Columbia University, New York, NY, USA;

[3] Materials Science & Technology Division, Naval Research Laboratory, Washington, DC, USA;

[4] Nova Research, Inc. Washington, DC, USA;

[5] Department of Aerospace and Mechanical Engineering, University of Notre Dame, Notre Dame, IN, USA;

[6] Horiba Scientific, Novato, CA, USA;

## Abstract

Second harmonic generation (SHG) is a nonlinear optical response arising exclusively from broken inversion symmetry in the electric-dipole limit. Recently, SHG has attracted widespread interest as a versatile and noninvasive tool for characterization of crystal symmetry and emerging ferroic or topological orders in quantum materials. However, conventional far-field optics is unable to probe local symmetry at the deep subwavelength scale. Here, we demonstrate near-field SHG imaging of 2D semiconductors and heterostructures with the spatial resolution down to 20 nm using a scattering-type nano-optical apparatus. We show that near-field SHG efficiency is greatly enhanced by excitons in atomically thin transition metal dichalcogenides. Furthermore, by correlating nonlinear and linear scattering-type nano-imaging, we resolve nanoscale variations of interlayer stacking order in bilayer $WSe_2$, and reveal the stacking-tuned excitonic light-matter-interactions. Our work demonstrates nonlinear optical interrogation of crystal symmetry and structure-property relationships at the nanometer length scales relevant to emerging properties in quantum materials.

**Introduction**

A wide range of emerging electronic, magnetic and topological properties in quantum materials are intimately linked to spatial inversion symmetry. In van der Waals heterostructures, the crystal symmetry can be precisely controlled by interlayer stacking and twisting techniques, providing a knob for programming the electronic bandstructure[1-3], magnetic orders[4], optical[5-11], thermal[12], phonon properties[13], as well as mechanical relaxation and atomic reconstruction[14]. Second harmonic generation (SHG) is the lowest-order nonlinear optical response, where the incident laser beam interacts with the material, emitting a frequency-doubled signal. The leading-order SHG arises from electric-dipole polarization, and is exclusively allowed in systems with a broken spatial inversion symmetry[15]. In addition to the crystallographic lattice structure, hidden orders of charge and spin configurations can also break inversion symmetry, such as ferroelectricity[16], antiferromagnetism[17-19], and even electron nematicity[20]. Therefore, SHG microscopy has grown in importance as a tool for directly accessing and visualizing crystal orientation and ferroic domain patterns. Since the spatial resolution attainable by far-field optics is limited by optical diffraction, developing scanning near-field SHG (*i.e.* "nano-SHG") imaging is highly desirable for investigation of symmetry properties at the nano- and meso-scale. This is the critical length scale relevant to many emerging quantum material systems, such as the lateral size of moiré superlattice[21, 22], van der Waals multiferroic domain patterns[23], and strain-localized 2D quantum dots[24].

Development of scanning near-field SHG imaging has proven to be a formidable task. Specifically, nano-SHG imaging of atomically-thin material is challenging because only an extraordinarily small area of sample material is available under the tip for nanoscale frequency conversion. In previous work, the low efficiency of SHG, exacerbated by weak nano-optical interactions, has generally required a fundamental pump laser with high pulse fluence of about 10 mJ/cm$^2$ [25], exceeding the reported damage threshold of plasmonic nanostructures[26] and most 2D semiconductors including transition metal dichalcogenides (TMDs)[27]. Hampered by tip and sample degradation, the achieved spatial resolution for nano-SHG has typically been ~100 nm[25, 28-31], with limited studies reaching ~50 nm on nonlinear ferroelectric oxide films[32]. In fact, sub-100-nm-level spatial resolution has not been achieved in SHG studies of 2D materials to date. Additionally, previous experiments also suffered from large background signal, including both the far-field SHG response from the sample and the SHG generated by the apex of the scanning probe (*i.e.*, tip-SHG)[25, 32]. As the background SHG is at the same wavelength with the desired near-field signal, disentangling the contributions is difficult, often requiring complex post-processing of image data based on first-principle models[33].

In this work, we realize nano-SHG imaging of atomically-thin TMD semiconductors and heterostructures with a spatial resolution of 20 nm. The aforementioned efficiency bottleneck is overcome by two-photon pumping of the band nesting excitons, providing SHG enhancement over a relatively broad range of frequency. We also generate near-field SHG signals with significantly lower background as evidenced by approach curve data. As a proof-of-principle demonstration, we image bilayer WSe$_2$ grown by chemical vapor deposition (CVD), resolving crystal domains with different interlayer stacking orders, and reveal the effects of stacking symmetry on nano-optical nonlinear light scattering.

**Results and Discussion**

Figure 1a shows a schematic of exciton-enhanced nano-SHG. The plasmonic tip is an atomic force microscopy (AFM) probe coated with Ti/Au (10/150 nm) layers for plasmonic enhancement. The tip is tilted with respect to the sample plane's normal direction at an angle of 30º. The fundamental pump laser has a pulse width of 140 fs and repetition rate of 80 MHz. The incidence direction (*i.e.* *k*-vector direction) of the pump is perpendicular to the axial orientation of the tip, and the electric field is set as *p*-polarized (*i.e.* parallel to the tip axis) to effectively excite the tip plasmon. With this configuration, the tip generates a localized plasmonic hot spot with both in-plane and out-of-plane electric field components at the fundamental frequency of $\omega$. The back-reflected SHG is collected by the same objective (effective Numerical Aperture, NA = 0.42) used for focusing the pump. More details about tip fabrication and the measurement setup are included in the Method section and Supplementary Information Figure S1 and S2. The second harmonic nature of the collected signal is verified by examining spectra position and the quadratic dependence of its intensity on pump power (Figure S3).

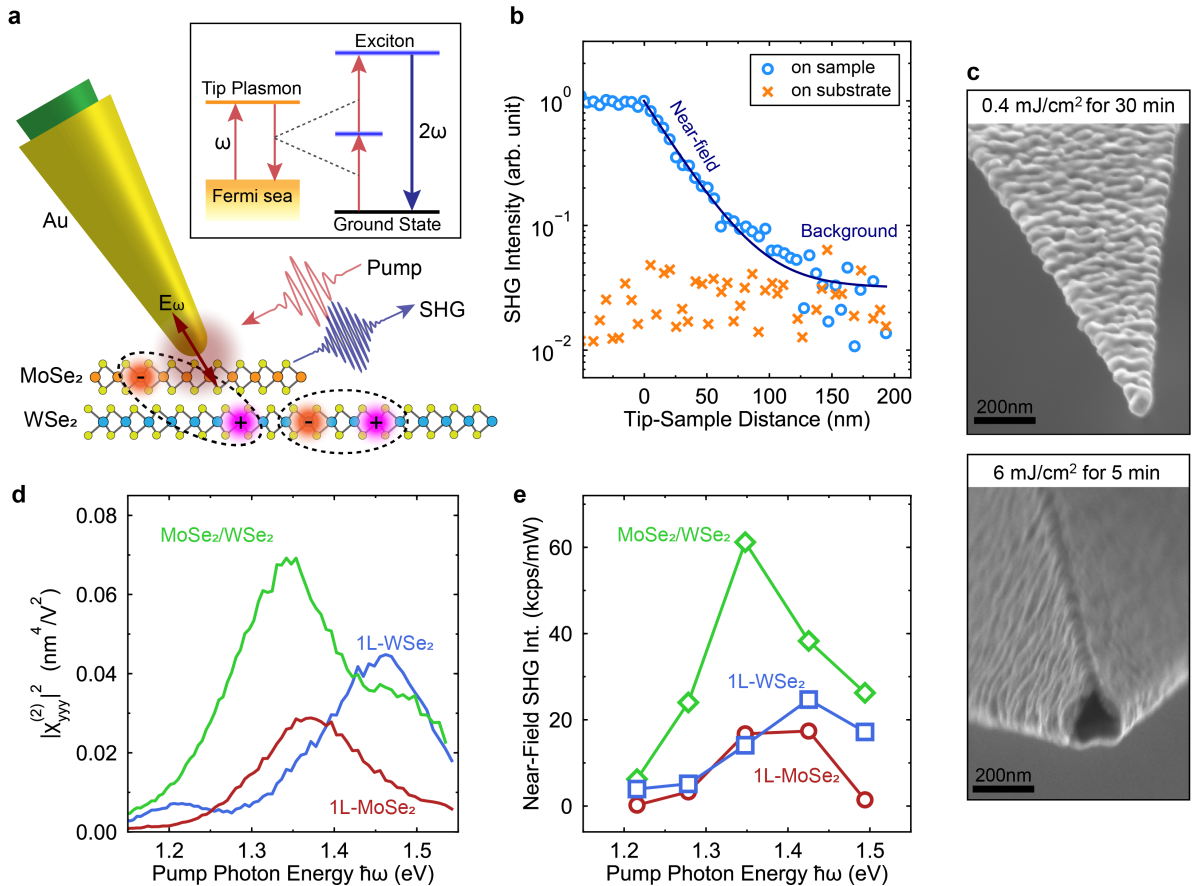

**Figure 1. Exciton-enhanced near-field second harmonic generation (SHG) from two-dimensional semiconductors and heterostructures**. (a) Schematic of near-field SHG (*i.e.* nano-SHG) from 2D semiconductors. The tip plasmon is excited by a femtosecond pump pulse at $\omega$, and SHG is enhanced by the exciton resonance at $2\omega$. (b) Approaching curves showing SHG intensity as a function of tip-sample distance. Solid blue line is a fit using exponential decay plus a flat background. (c) Scanning electron microscopy (SEM) of Au-coated plasmonic tips after scanning with low (upper) and high (lower) pump fluence. (d) Dependence of nonlinear susceptibility $|\chi^{(2)}|$ on pump photon energy as measured in the far-field, showing enhancement by band nesting excitons. (e) Dependence of near-field SHG efficiency on pump photon energy. kcps: kilo counts per second. mW: miliWatt.

To establish the near-field nature of SHG, we measure the dependence of SHG intensity as a function of tip-sample distance on a monolayer $WSe_2$ sample (Figure 1b). SHG intensity follows an exponential decay as the tip is retracted away from the sample, providing compelling evidence for the evanescent-wave nature of the tip-localized plasmonic hot spot. At further distance, slight deviation from the exponential trend is observed, which is attributed to a weak residual far-field background. We fit the approach curve data with contributions from the near-field (a single exponential decay) and the background (a flat baseline), as shown by the solid line in Figure 1b. The 1/e exponential decay length is found to be 23±3 nm. Analogous to tip-enhanced Raman and photoluminescence[34, 35], the near-field enhancement factor for nano-SHG can be defined as the ratio of near-field signal over the background. The average enhancement factor is deduced to be 18±7 (see Supplementary Figure S4 for more details), indicating that the collected signal is dominated by local response within the evanescent field of the tip, with minimal interference from far-field SHG or tip-SHG background. Therefore, no post-processing is required for data interpretation. With the large pump incident angle (~60°), the reflected far-field SHG will primarily propagate in the forward direction (towards the left in Figure 1a) due to lateral phase matching rather than back into the collection objective.

To achieve the highest possible spatial resolution, we find it critical to avoid ablation damage of the tip apex. Therefore, the pump fluence should be well controlled. Previous work reported that with the femtosecond laser, damage threshold of Au plasmonic nanostructure is on the order of 1 $mJ/cm^2$ [26]. In our experiments, when the pulse fluence is set at 0.4 $mJ/cm^2$, the coated metal layer remains continuous and conformally covers the entire apex with Au grains, as shown by scanning electron microscopy (SEM) images in Figure 1c, upper panel. In comparison, when the pulse fluence is increased to 6 $mJ/cm^2$, the plasmonic apex is severely damaged after only 5 minutes of exposure (Fig. 1c, lower panel). Additional SEM images are included in the Supplementary Figure S5.

The second order nonlinear susceptibilities of atomically thin TMDs show large enhancement on exciton resonances in the visible and near-infrared range[36]. Figure 1d shows the far-field SHG efficiency measured for representative monolayer (1L-) $WSe_2$, 1L-$MoSe_2$, and 3R-stacked $MoSe_2/WSe_2$ heterobilayer samples used in this work. Here we quantify the SHG efficiency as the square of sheet nonlinear susceptibility, $|\chi^{(2)}|^2$, which is proportional to SHG intensity under normalized pump power. The spectra are highly dispersive with peaks corresponding to two-photon resonances of band nesting excitons in the respective materials[36]. Figure 1e shows the near-field SHG efficiency at different pump photon energies, quantified as kilocounts per second (kcps) per miliwatt (mW) reaching the first collection lens. Exciton enhancement is also pronounced in the near-field data, with nano-SHG efficiency improving by an order of magnitude when the pump photon energy is in two-photon resonance with band nesting excitons.

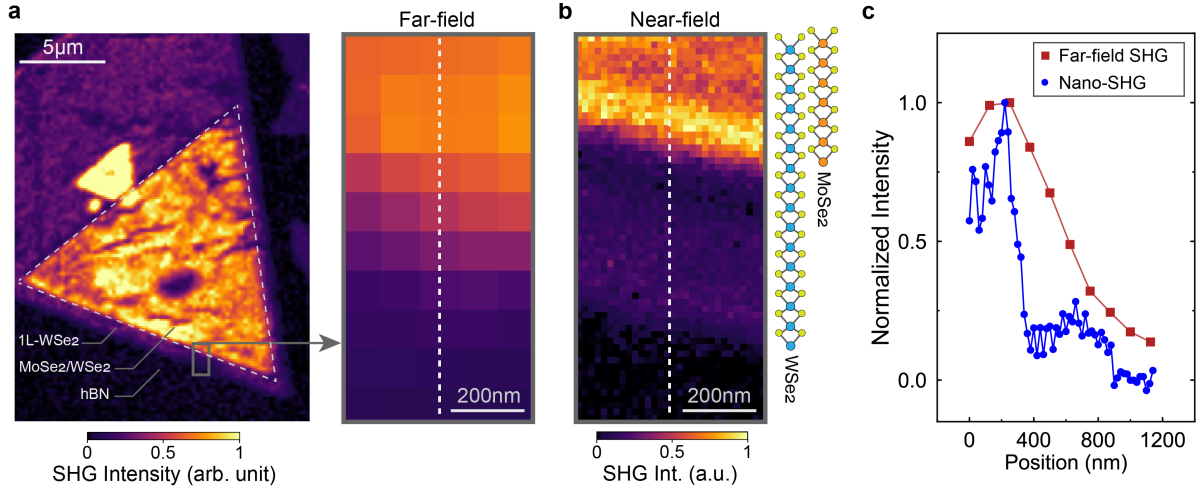

**Figure 2. Nano-SHG imaging of a 2D semiconductor heterostructure with sub-diffraction-limited spatial resolution.** (a) Far-field confocal SHG imaging of a 3R-stacked $MoSe_2/WSe_2$ heterobilayer on hBN substrate. The heterobilayer region is enclosed by the white dashed triangle. The right panel shows a zoom-in view of the boxed region. (b) Nano-SHG imaging of the same boxed region as in (a), with schematic illustration of the $MoSe_2$ and $WSe_2$ layers on the right. (c) Comparing far-field and nano-SHG intensities along the dashed white lines in (a, b).

Nano-SHG scanning probe imaging of a $MoSe_2/WSe_2$ heterobilayer sample is shown in Figure 2. The sample is created by transferring a CVD-grown triangular monolayer $MoSe_2$ flake onto a larger triangular $WSe_2$ flake, with 3R-type interlayer stacking order. The TMD layers are on a hexagonal boron nitride (hBN) substrate ~20 nm thick. The entire van der Waals stack is on a $SiO_2$/Si substrate. Figure 2a shows the far-field SHG microscopy. The overlapped $MoSe_2/WSe_2$ heterobilayer area is bounded by the white dashed triangle. The brightest small triangle on the top-left side of the heterobilayer is a nucleation region with 3R-stacked bilayer $WSe_2$, and the stacking effects on SHG is discussed in latter part of this work. The $MoSe_2/WSe_2$ heterobilayer shows strong SHG intensity compared to 1L-$WSe_2$, consistent with measurements of the nonlinear susceptibility shown in Figure 1d and 1e. As SHG intensity from the hBN substrate is highly thickness and pump wavelength dependent[5], the hBN substrate used here is properly chosen so that the substrate SHG intensity is more than one order of magnitude lower than that from 1L-$WSe_2$. The right panel of Figure 2a shows a zoom-in view of the confocal SHG image in the boxed region where, from top to bottom, the material changes from $MoSe_2/WSe_2$ heterobilayer to 1L-$WSe_2$ and finally to the hBN substrate. The intensity transitions within a sub-micrometer length scale and thus appears blurred in the far-field image due to diffraction-limited resolution.

Figure 2b shows the nano-SHG image of the same boxed region. The image has 30 by 60 pixels with an integration time of one second per pixel, and photon energy of the fundamental pump is set at 1.27 eV. In this case, the sharp SHG contrast transition across these materials is clearly resolved, demonstrating the nanoscale spatial resolution of SHG. Figure 2c compares the SHG intensity profiles along the linecuts in far- and near-field images. A 20 nm sharp step transition of SHG intensity is observed from 1L-WSe$_2$ to the hBN substrate, which is estimated as the spatial resolution. The resolution agrees well with the radius of curvature of tip apex seen in SEM images, and matches the exponential decay length of evanescent wave extracted from the approach curve data in Figure 1b. In addition, the near-field SHG intensity is enhanced along the edge of the MoSe$_2$ layer. Edge-enhanced SHG has previously been observed by far-field SHG microscopy, showing bright edges with full width half magnitude (FWHM) on the order of several hundreds of nm, limited by diffraction[37,38]. It was proposed that atomic termination and reconstruction along the crystal edge may introduce one-dimensional midgap states, causing resonant enhancement of a nonlinear optical transition. In the nano-SHG images shown here, we find the bright edge region has a FWHM of ~70 nm and ~110 nm for the samples presented in Figure 2b and Figure 3c, respectively. Both are significantly wider than the near-field spatial resolution (~20 nm). Therefore, the edge enhancement in this work occurs within a finite-width ribbon-like region, rather than an atomic-scale 1D edge state. Our results suggest that edge-related mesoscale disorders, such as increased defect density[39], could also play a role in edge-enhanced nonlinear response.

Next we demonstrate that local interlayer stacking order can be identified and visualized by nano-SHG. Figure 3a shows an optical microscope image of a CVD grown WSe$_2$ sample transferred onto SiO$_2$ substrate. The flake of interest includes both monolayer and bilayer regions. The dark-colored bilayer region has a jigsaw-like pattern, which is formed by multiple triangular domains with 3R and 2H stacking order[40], as confirmed by following nonlinear nano-SHG and linear s-SNOM scattering. AFM topography of the boxed area is shown in Figure 3b, where the bilayer and monolayer regions can be clearly distinguished by height. No appreciable topographic contrast is observed within the bilayer region. To investigate the stacking orders of these bilayer domains, the nano-SHG image is acquired from the same region, as shown by Figure 3c. This image is 40 by 40 pixels in size with an integration time of 0.4 seconds per pixel, and photon energy of the fundamental pump is 1.41 eV. In stark contrast to the topographic image showing a homogeneous bilayer, a prominent SHG contrast is observed. The upper bright region corresponds to a 3R-stacked bilayer with broken inversion symmetry, where the nonlinear dipole polarization from top and bottom layers are in-phase, thus resulting in the constructive SHG response. The 3R-stacking is further confirmed by polarization-resolved far-field SHG measurements. In contrast, no SHG is detected from the bottom-left region, consistent with the 2H-stacked bilayer structure that preserves inversion symmetry.

Quantitatively, the collected nano-SHG intensity from the 3R-bilayer and the monolayer have average values of 620 cps and 160 cps, respectively, with the intensity ratio being close to four. The intensity ratio indicates that nano-SHG response is dominated by in-plane dipole moments as sketched by Figure 3c, where the 3R-bilayer has aligned dipole moments with constructive interference, giving four times the intensity of a monolayer. Out-of-plane $\chi^{(2)}$ components ($\chi^{(2)}_{xyz}$, $\chi^{(2)}_{xzy}$, $\chi^{(2)}_{zxy}$) are allowed in principle based on the point group symmetry (D$_3$) of 3R-homobilayer TMD. However, if the out-of-plane contribution was significant, we would expect the intensity ratio to be significantly higher than four, since the additional out-of-plane $\chi^{(2)}$ components are absent in a monolayer. We also find that nano-SHG intensity from the centrosymmetric 2H-bilayer is negligible (well below our detection noise floor of ~20 cps). This finding confirms that the presence of the scanning probe does not significantly alter or break local symmetry, or at least is insignificant compared to the intrinsic crystal stacking order.

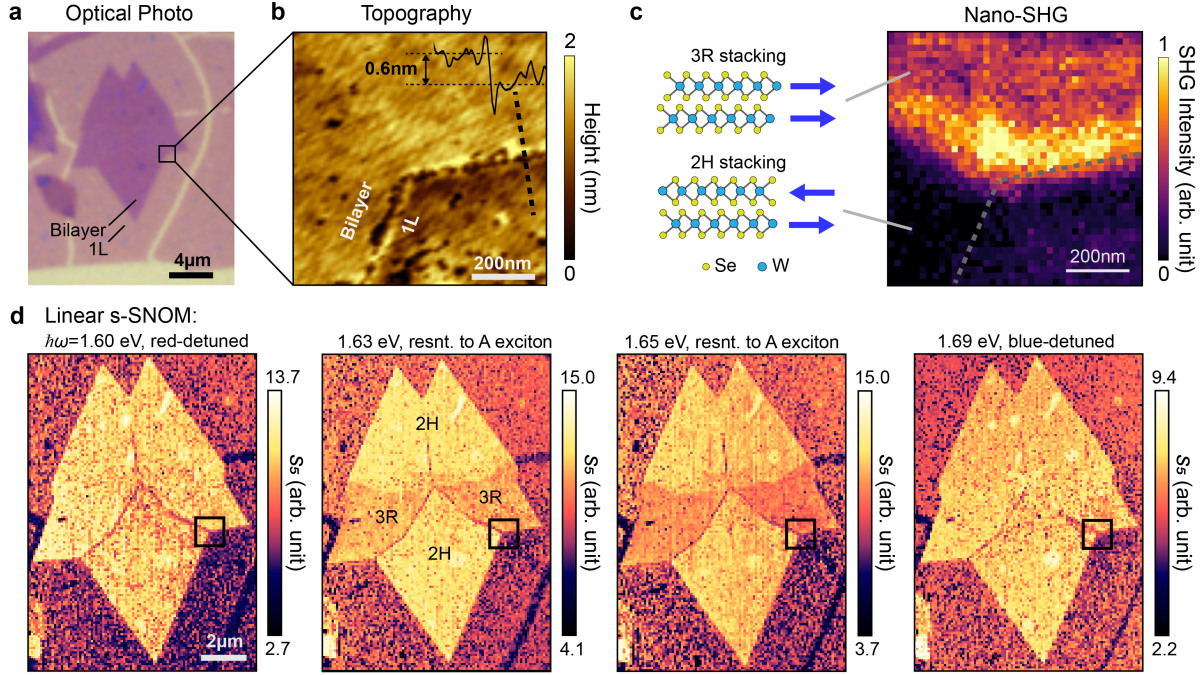

**Figure 3. Revealing local interlayer stacking order and stacking effects on excitonic light-matter-interaction.** (a) Optical microscope photo of the WSe$_2$ sample with bilayer and monolayer (1L) regions. (b) AFM topography and (c) nano-SHG image of the boxed area in (a). Height profile along the black dashed line in (b) reveals the monolayer-to-bilayer step. The parallel (anti-parallel) interlayer lattice alignment for 3R(2H)-bilayer domains are shown by (c), leading to constructive (destructive) interference of local SHG response. (d) Images of linear scattering-SNOM amplitude acquired with the incident photon energies on- and off-resonance with the A excitons. Clear contrast between 3R and 2H stacking is observed on exciton resonance. Boxed regions in (d) correspond to the nano-SHG scanning area in (c). $S_5$: fifth-harmonic of scattering amplitude.

With the 3R and 2H crystal domains being visualized, we further investigate the effect of different stacking symmetry on light-matter interactions by performing linear optical scattering SNOM measurements. The scattered signal is governed by the dielectric function of a material, thus allowing one to investigate the exciton resonance energy and oscillator strength[41]. Figure 3d shows the s-SNOM images of the scattering amplitude demodulated at the fifth harmonic of tip tapping frequency, $S_5$, which is proportional to local reflectance contrast within the weak resonance limit[41]. A range of incident photon energies are applied around the A exciton resonance of bilayer WSe$_2$ which is expected to be at around 1.65 eV. Recent far-field spectroscopy showed variation of optical reflectance between 2H and 3R bilayers[40]. Here in the near-field we observed wavelength-dependent contrast in nano-optical light scattering across the crystal domains. When the photon energy is either red-detuned (1.60 eV) or blue-detuned (1.69 eV) from the A excitons, the entire bilayer flake shows uniform scattering amplitude, suggesting that the stacking effect on the optical dielectric function is rather weak in the non-resonant frequency range. However, when the incident laser resonantly excited the A exciton (1.63 eV and 1.65 eV data), contrast between the domains clearly emerges[40]. In accordance to previous theoretical calculations[42], the breaking of inversion symmetry in the 3R-bilayer leads to additional splitting of the valance band maximum around K points, which further results in split exciton resonances. Thus, correlating these complementary nonlinear optical imaging modalities allows us to unambiguously establish the crystal-symmetry-based origins of the contrast in local optical response.

**Outlook**

In summary, we report nano-SHG imaging of atomically-thin transition metal dichalcogenides and their heterostructures, and demonstrate its application for nanoscale visualization of interlayer crystal stacking orders. Our strategy of exploiting the strong exciton enhancement of nonlinear optical response to boost nano-SHG efficiency and perform scanning probe imaging enables an order of magnitude lower pump fluence compared to non-resonant cases, avoiding laser ablation damage to the plasmonic tip. Further, we demonstrate the power of combining nonlinear and linear nano-optics for correlating local variations of atomic registration and interlayer stacking to changes in excitonic light-matter-interaction, revealing intimate structure-property relationships at the length scales relevant to quantum materials. Since SHG is sensitive to a variety of symmetry-breaking orders, we expect that the nano-SHG approach highlighted by these results can be widely applied to visualize subwavelength domains of crystal structure, ferroelectric and magnetic orders. More generally, it should be applicable to the expansive collection of novel heterostructures exhibiting inherent and programmably broken symmetries[5], whose exceptional properties have yet to be fully unlocked.

**Acknowledgements**

The authors thank S. Moore for valuable discussion. This work is supported by Programmable Quantum Materials, an Energy Frontier Research Center funded by the US Department of Energy (DOE), Office of Science, Basic Energy Sciences (BES), under award DE-SC0019443. Work at the Naval Research Lab was supported by internal core programs.

**Methods**

*Fabrication of plasmonic scanning probes for nano-SHG.* We start with commercial silicon atomic force microscopy (AFM) probes with a resonance frequency of 300 (or 60) kHz and spring constant of 100 (or 2.7) N/m. The stiffer cantilevers are more stable in scanning probe operation with the pulsed laser, but in principle both kinds of AFM probes can work. Layers of Ti/Au are coated by electron beam evaporation, with a rate of 2 angstrom/s in a chamber with base pressure of $10^{-8}$ to $10^{-7}$ Torr. The coating thickness of Ti/Au is 2/30 nm. Note that since the AFM tip is mounted with the apex facing towards the crucible, we estimate that the actual coating thickness on the sidewall of the pyramid-shaped shaft is about 1/5 of the nominal values. So the evaporation thickness of Ti/Au as registered by the crystal monitor is set to be 10/150 nm. Tips coated with different thicknesses of Au have been tested, and 150 nm is found to provide low laser ablation damage and negligible SHG from the tip itself at our typical power levels for nano-SHG imaging.

*Nano-SHG measurement setup and procedure.* The nano-SHG setup is home-built based on an atomic force microscope (Omegascope, Horiba Inc.). Schematics of the full setup is shown in Supplementary Figure S1. The fundamental pump laser is a Ti:Saph laser (Coherent) with 140 fs pulse width and 80 MHz repetition rate. The objective lens is 100X with effective (beam is partially blocked by substrate) numerical aperture of 0.42. Sample is placed on a tilted wedge for controlling tip angle. The collected signal is passed through appropriate spectral filters, a 100 µm pin hole, and dispersed by a 150 g/mm holographic grating onto an electron magnifying charge coupled device (EMCCD, Horiba Synapse FIVS). The system collection efficiencies are determined to be 2.8%, 4.4%, 7.9%, 13.4%, 18.1% for wavelengths at 415 nm, 435 nm, 460 nm, 485 nm, 510 nm, respectively, which are used for calculating nano-SHG efficiency reaching the first lens (in Figure 1e). In nano-SHG imaging, the AFM operates under tapping mode when going from one pixel to the next. At each pixel, the tip goes into contact with the sample for a set integration time during which the SHG is measured.